# Specific Heat and Thermal Entanglement in an Open Quantum system

B. Lari[1,*], and H. Hassanabadi[2]

[1] Department of Physics, Ahvaz Branch, Islamic Azad University, Ahvaz, Iran
[2] Physics Department, Shahrood University of Technology, Shahrood, Iran, P. O. Box 3619995161-316

E-mail: [1] behzadlari1979@yahoo.com[*] and   [2] hha1349@gmail.com

## Abstract

In this paper, the density matrix is obtained using the Non-Markovian master equation method for a system consisting of two qubit modeling with the Heisenberg XXZ chain, which involves the Dzyaloshinskii–Moriya interaction and exposed to the bosonic baths. Also, using the proposed formula for calculating the specific heat through the density matrix of the open quantum system, the specific heat is calculated. The quantum entanglement behavior with time and coupling constant is investigated. It is observed that, the specific heat at low temperatures becomes negative when the system is exposed to the environment. The time behavior of quantum entanglement for this model showed that the entanglement decreases to zero with increasing time.

## 1. Introduction

To fabricate quantum gates and memories [1-4], the quantum entanglement of the system and proper choice of the material according to environmental parameters is of great importance. The engaged parameters in the studies include the specific heat and the constant magnetic coupling [5]. In this regards, some researchers have tried to find a relationship between the heat capacity and the maximum quantum entanglement for closed systems [6-7]. Others considered the entanglement area law from specific heat capacity and derived the relation between exponential decay of specific heat capacity at low temperature and the entanglement at low-energy state [8]. In this paper we consider one of the standard models, namely, the anisotropic Heisenberg XXZ regime for two qubit system with the Dzyaloshinskii–Moriya (DM) interaction [9-10]. The model arises from the spin–orbit coupling where each of the qubits are exposed to separate bosonic baths. Many scientific reports have investigated the entanglement [11-15] and its specific heat [16-17] in different open quantum systems. Nevertheless, it seems useful to present a formula to calculate the specific heat (as a determinative thermodynamics parameter of solids) using density matrix and its eigenvalues for an open quantum system. In order to calculate the time evolution of the density matrix of system, we use a unitary transform to the interaction picture and solve the Von-Neumann equation with the Born-approximation [18-19]. It is assumed that, the bosonic baths obey the Cauchy-Lorentz distribution of frequencies. The time evolution of entanglement with the effective parameters in the Hamiltonian of system, such as spin-orbit coupling $D_z$ and coupling constant $J_i$ (where $i \in \{x,y,z\}$), is useful in practical fabrication of graphene-based (or GaAs-based) quantum gates [5], [20-25]. Therefore, in order to achieve our goals, this paper has been organized as follows. In Sec. 2, using the Hamiltonian of anisotropic Heisenberg XXZ model for two qubits with the DM Interaction exposed to separate bosonic baths, we obtain the initial density matrix of the system as a function of temperature and solve the Non-Markovian Master equation for our model. In Sec. 3, we have proposed a formula to calculate the specific heat using density matrix of the open quantum system. Also, the entanglement of formation (EOF) is computed. In Sec. 4, the obtained results will be summarized.

## 2. Non-Markovian Approach

The Hamiltonian of anisotropic Heisenberg XXZ model for two-qubit system with D-M interaction, such that each qubit be exposed to Bosonic bath, can be defined as follows,

$$H = H_s + H_b + H_{sb} \quad (1)$$

In which

$$H_s = \sum_i J_i \, \sigma_1^i \sigma_2^i + D_z\big(\sigma_1^x \sigma_2^y - \sigma_1^y \sigma_2^x\big) \quad , i \in \{x,y,z\}) \quad (2)$$

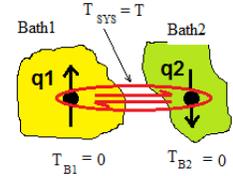

$$H_b = \sum_{n_1} \omega_{n_1} b_{n_1}^\dagger b_{n_1} + \sum_{n_2} \omega_{n_2} b_{n_2}^\dagger b_{n_2} \quad (3)$$

$$H_{sb} = \sum_{i=1}^{2} \sigma_i^+ \sum_{n_i} g_{n_i} b_{n_i} + h.c. \quad (4)$$

$H_s$ is Hamiltonian of the system with DM interaction. $H_b$ shows the Hamiltonian of baths including $n_i$ mode with frequency $\omega_{n_i}$. $H_{sb}$ denotes the Hamiltonian of the interaction between the system and the baths with strength interaction $g_{n_i}$. "$i$" labels the first and the second qubit and $\sigma_{\ldots}^i$ is the Pauli matrix. The coupling constant $J_i > 0$ corresponds to the antiferromagnetic case and $J_i < 0$ corresponds to the ferromagnetic case. The model is called Heisenberg XXZ model if $J_x = J_y = J$ and $J_z \neq J$. The density matrix of system $\rho_s(T)$, is obtained in terms of the standard basis $\{|00\rangle, |10\rangle, |01\rangle, |11\rangle\}$ as

$$\rho_s(t=0,T) = \frac{e^{-\beta H_s}}{Z}$$

$$= \frac{1}{Z}\{e^{-\beta J_z}|00\rangle\langle 00| + u|10\rangle\langle 10| + v e^{i\theta}|01\rangle\langle 10| + v e^{-i\theta}|10\rangle\langle 01| + u|01\rangle\langle 01| + e^{-\beta J_z}|11\rangle\langle 11|\} \quad (5)$$

where $\beta = \frac{1}{k_B T}$ ($k_B$ is Boltzmann constant) and

$$u = \frac{1}{2} e^{\beta(J_z - 2\eta)}(1 + e^{4\beta\eta})$$

$$v = \frac{1}{2} e^{-\beta(J_z - 2\eta)}(1 - e^{4\beta\eta})$$

$$Z = 2\ e^{-\beta J_z}[1 + e^{2\beta J_z} \cosh(2\beta\eta)] \quad (6)$$

$$\eta = \sqrt{J^2 + D_z^2} \ ;\ \theta = \tan^{-1}\left(\frac{D_z}{J}\right)$$

The density matrix defined in Eq. (5), has the X-type form [26]. We start with Von-Liouville equation and find the time evolution of density matrix

$$\frac{\partial \rho_{sb}(t)}{\partial t} = -\frac{i}{\hbar}[H(t), \rho_{sb}(t)] \quad (7)$$

In the rest of the paper, we set $\hbar = 1$. To obtain the Non-Markovian master equation we apply the Born-approximation

$$\frac{\partial \rho_s^I(t)}{\partial t} = -i\ tr_b\{[H_{sb}^I(t), \rho_{sb}^I(0)]\} - \int_0^t d\acute{t}\ tr_b\{[H_{sb}^I(t), [H_{sb}^I(\acute{t}), \rho_{sb}^I(\acute{t})]]\} \quad (8)$$

Where the superscript "$I$", denotes the interaction picture. The first term in Eq. (8) is considered as zero (see the Appendix and [27-28]).

Therefore, we have

$$\frac{\partial \rho_s^I(t)}{\partial t} = -\int_0^t d\acute{t}\ tr_b\{[H_{sb}^I(t), [H_{sb}^I(\acute{t}), \rho_s^I(t) \otimes \rho_b^I(0)]]\} \quad (9)$$

where
$$H_{sb}^I(t) = \exp\{i(H_s + H_b)t\} H_{sb} \exp\{-i(H_s + H_b)t\} \quad (10)$$

$$\rho_s^I(t) = \exp\{iH_s t\}\rho_s(t)\exp\{-iH_s t\} \quad (11)$$

$$\rho_b^I(0) = \rho_b(0) \quad (12)$$

We rewrite Eq. (4) in interaction picture as

$$H_{sb}^I = \sum_{j=1}^{2} \sigma_j^+ \sum_{n_j} g_{n_j} b_{n_j} \exp\{-i\omega_{n_j} t\} + h.c. \quad (13)$$

To derive the above equation, we have used the following relations

$$\sigma_j^+(t) = Tr_i\left\{\exp\left(\frac{iH_s t}{\hbar}\right)\sigma_j^+(0)\exp\left(\frac{-iH_s t}{\hbar}\right)\right\}$$

$$for\ i,j \in \{1,2\}\ and\ i \neq j \Rightarrow \sigma_j^+(t) = \sigma_j^+(0) = \sigma_j^+$$

$$\frac{db_{\acute{n}_j}(t)}{dt} = i\left[H_b, b_{\acute{n}_j}(t)\right] \Rightarrow b_{n_j}(t) = b_{n_j}\exp\{-i\omega_{n_j} t\}$$

Apart from the physical dimensions, we can choose the following initial state to the baths.

$$\rho_b(0) = |N;\ldots,n_j,\ldots,n_p,\ldots\rangle_b \langle N;\ldots,n_j,\ldots,n_p,\ldots|$$

Finally, after tracing over the baths and using the above definition for $\rho_b(0)$, we have

$$\frac{d\rho_s^I(t)}{dt} = \sum_{j=1}^{2} \{ [\sigma_j^- \rho_s^I(t), \sigma_j^+] \sum_{n_j} |g_{n_j}|^2 n_j \int_0^t d\acute{t}\, e^{-i\omega_{n_j}(t-\acute{t})}$$

$$+ [\sigma_j^+, \rho_s^I(t)\sigma_j^-] \sum_{n_j} |g_{n_j}|^2 (n_j+1) \int_0^t d\acute{t}\, e^{-i\omega_{n_j}(t-\acute{t})} + h.c. \}$$

$$= \sum_{j=1}^{2} \mathcal{T}_j(t) \rho_s^I(t) \qquad (14)$$

If we suppose the bath is fixed on zero temperature, then the initial state has the form $|N; 0, \ldots, n_i = 0, \ldots, 0\rangle_b$. Therefore, the first term in Eq. (14) is neglected. Finally we have

$$\frac{d\rho_s^I(t)}{dt} = \sum_{j=1}^{2} \{ [\sigma_j^+, \rho_s^I(t)\sigma_j^-] \sum_{n_j} |g_{n_j}|^2 \int_0^t d\acute{t}\, e^{-i\omega_{n_j}(t-\acute{t})} + h.c. \} = \sum_{j=1}^{2} \mathcal{T}_j(t) \rho_s^I(t) \qquad (15)$$

$\mathcal{T}_j(t)$ is a super operator. If we suppose the baths are bosonic systems with Cauchy-Lorentz distribution, we have

$$\frac{d\rho_s^I(t)}{dt} = R(t) \sum_{j=1}^{2} [\sigma_j^+, \rho_s^I(t)\sigma_j^-] + h.c. = \sum_{j=1}^{2} \mathcal{T}_j(t) \rho_s^I(t) \qquad (16)$$

With $$R(t) = \int_0^t d\acute{t} \int_{-\infty}^{+\infty} d\omega\, J(\omega)\, e^{i\omega \acute{t}} = \frac{\gamma_0}{2}(1 - e^{-\gamma t})$$

The Markovian limit can be constructed by taking the limit $t \to \infty$. To derive $R(t)$, one can substitute the $\sum_{n_j} |g_{n_j}|^2$ by $\int_{-\infty}^{+\infty} d\omega\, J(\omega)$ and then use the relation

$$J(\omega) = \frac{\gamma_0}{\pi} [\frac{\gamma^2}{\omega^2 + \gamma^2}]$$

Where $\gamma$ is the scale parameter which specifies the half-width at half-maximum. Finally, transforming back to the Schrodinger picture we obtain the master equation

$$\frac{d\rho_s(t)}{dt} = -i[H_s, \rho_s(t)] + \sum_{j=1}^{2} \mathcal{T}_j(t)\, \rho_s(t) \qquad (17)$$

After a little algebra, we can obtain three independent differential equations on components of $\rho_{ij}(t)$:

$$\frac{d}{dt}\begin{pmatrix}\rho_{14}(t,T)\\ \rho_{41}(t,T)\end{pmatrix} = \begin{pmatrix}-\gamma_0(1-e^{-\gamma t}) & 0 \\ 0 & -\gamma_0(1-e^{-\gamma t})\end{pmatrix}\begin{pmatrix}\rho_{14}(t,T)\\ \rho_{41}(t,T)\end{pmatrix} \qquad (18)$$

$$\frac{d}{dt}\begin{pmatrix}\rho_{12}(t,T)\\ \rho_{13}(t,T)\\ \rho_{24}(t,T)\\ \rho_{34}(t,T)\end{pmatrix} = \begin{pmatrix}-2iJ_z - \frac{3\gamma_0}{2}(1-e^{-\gamma t}) & i(2J - 2iD_z) & 0 & 0 \\ i(2J + 2iD_z) & -2iJ_z - \frac{3\gamma_0}{2}(1-e^{-\gamma t}) & 0 & 0 \\ 0 & \gamma_0(1-e^{-\gamma t}) & 2iJ_z - \frac{\gamma_0}{2}(1-e^{-\gamma t}) & -i(2J + 2iD_z) \\ \gamma_0(1-e^{-\gamma t}) & 0 & -i(2J - 2iD_z) & -2iJ_z - \frac{\gamma_0}{2}(1-e^{-\gamma t})\end{pmatrix}\begin{pmatrix}\rho_{12}(t,T)\\ \rho_{13}(t,T)\\ \rho_{24}(t,T)\\ \rho_{34}(t,T)\end{pmatrix} \qquad (19)$$

$$\frac{d}{dt}\begin{pmatrix}\rho_{11}(t,T)\\ \rho_{22}(t,T)\\ \rho_{33}(t,T)\\ \rho_{44}(t,T)\\ \rho_{23}(t,T)\\ \rho_{32}(t,T)\end{pmatrix}=\begin{pmatrix}-\frac{4\gamma_0}{2}(1-e^{-\gamma t}) & 0 & 0 & 0 & 0 & 0\\ \gamma_0(1-e^{-\gamma t}) & -\gamma_0(1-e^{-\gamma t}) & 0 & 0 & i(2J-2iD_z) & -i(2J+2iD_z)\\ \gamma_0(1-e^{-\gamma t}) & 0 & -\gamma_0(1-e^{-\gamma t}) & 0 & -i(2J-2iD_z) & i(2J-2iD_z)\\ 0 & \gamma_0(1-e^{-\gamma t}) & \gamma_0(1-e^{-\gamma t}) & 0 & 0 & 0\\ 0 & i(2J+2iD_z) & -i(2J+2iD_z) & 0 & -\gamma_0(1-e^{-\gamma t}) & 0\\ 0 & -i(2J-2iD_z) & i(2J-2iD_z) & 0 & 0 & -\gamma_0(1-e^{-\gamma t})\end{pmatrix}\begin{pmatrix}\rho_{11}(t,T)\\ \rho_{22}(t,T)\\ \rho_{33}(t,T)\\ \rho_{44}(t,T)\\ \rho_{23}(t,T)\\ \rho_{32}(t,T)\end{pmatrix} \quad (20)$$

We can rewrite the above equations in as

$$\frac{d}{dt}|\alpha(t)\rangle_i = \widehat{M}_i(t)|\alpha(t)\rangle_i, \quad i \in \{1,2,3\} \quad (21)$$

The "i" labels each of the above differential equations. Using the initial state of the system defined in Eq. (5), we can see, all elements of density matrix are zero unless those ($i = 3$) defined in Eq. (20). To solve the Eq. (21), we use the following similarity transformation P

$$T e^{\int_0^t \widehat{M}_i(\hat{t})d\hat{t}} = P\, e^G P^{-1}$$

where "G" is called the Jordan form of $\int_0^t \widehat{M}_i(\hat{t})d\hat{t}$ and the symbol $T$ denotes the standard time ordering in the exponent. The matrix $\widehat{M}_3$ can be rewritten as

$$\widehat{M}_3(t) = \hat{A} + \gamma_0 e^{-\gamma t}\hat{B}$$

where $\hat{A}$ and $\hat{B}$ are the constant matrices with no contribution in the time ordering.

$$\hat{A} = \begin{pmatrix}-4 & 0 & 0 & 0 & 0 & 0\\ 2 & -2 & 0 & 0 & 0 & 0\\ 2 & 0 & -2 & 0 & 0 & 0\\ 0 & 2 & 2 & 0 & 0 & 0\\ 0 & 0 & 0 & 0 & -2 & 0\\ 0 & 0 & 0 & 0 & 0 & -2\end{pmatrix}; \quad \hat{B} = \begin{pmatrix}0 & 0 & 0 & 0 & 0 & 0\\ 0 & 0 & 0 & 0 & 2D_z & 2D_z\\ 0 & 0 & 0 & 0 & -2D_z & 2D_z\\ 0 & 0 & 0 & 0 & 0 & 0\\ 0 & w & -w & 0 & 0 & 0\\ 0 & w^* & -w^* & 0 & 0 & 0\end{pmatrix}$$

Therefore, we have the commutation relation for $\widehat{M}_3$

$$[\widehat{M}_3(t_1), \widehat{M}_3(t_2)] = 0$$

and we can write

$$e^{\left(\left(\int_0^t d\hat{t}\right)\hat{A}+\left(\gamma_0 \int_0^t e^{-\gamma \hat{t}}d\hat{t}\right)\hat{B}\right)} = P\, e^G P^{-1}$$

The general solution for Eq. (21), is

$$|\alpha(t)\rangle_i = P\, e^G P^{-1}|\alpha(0)\rangle_i \quad (22)$$

As a result, we have the following relation to calculate non-zero elements of density matrix

$$\begin{pmatrix}\rho_{11}(t,T)\\\rho_{22}(t,T)\\\rho_{33}(t,T)\\\rho_{44}(t,T)\\\rho_{23}(t,T)\\\rho_{32}(t,T)\end{pmatrix}=$$

$$\begin{pmatrix} e^{-4B} & 0 & 0 & 0 & 0 & 0 \\ -e^{-4B}+e^{-2B} & \frac{e^{-2B}}{2}(1+\cos[2|\Gamma|.t]) & \frac{e^{-2B}}{2}(1-\cos[2|\Gamma|.t]) & 0 & \frac{e^{-2B}}{2}\frac{|\Gamma|}{\Gamma}\sin[2|\Gamma|.t] & \frac{e^{-2B}}{2}\frac{|\Gamma|}{\Gamma^*}\sin[2|\Gamma|.t] \\ -e^{-4B}+e^{-2B} & \frac{e^{-2B}}{2}(1-\cos[2|\Gamma|.t]) & \frac{e^{-2B}}{2}(1+\cos[2|\Gamma|.t]) & 0 & -\frac{e^{-2B}}{2}\frac{|\Gamma|}{\Gamma}\sin[2|\Gamma|.t] & -\frac{e^{-2B}}{2}\frac{|\Gamma|}{\Gamma^*}\sin[2|\Gamma|.t] \\ 1+e^{-4B}-2e^{-2B} & 1-e^{-2B} & 1-e^{-2B} & 1 & 0 & 0 \\ 0 & -\frac{e^{-2B}}{2}\frac{|\Gamma|}{\Gamma^*}\sin[2|\Gamma|.t] & \frac{e^{-2B}}{2}\frac{|\Gamma|}{\Gamma^*}\sin[2|\Gamma|.t] & 0 & \frac{e^{-2B}}{2}(1+\cos[2|\Gamma|.t]) & -\frac{\Gamma}{\Gamma^*}\frac{e^{-2B}}{2}(1-\cos[2|\Gamma|.t]) \\ 0 & -\frac{e^{-2B}}{2}\frac{|\Gamma|}{\Gamma}\sin[2|\Gamma|.t] & \frac{e^{-2B}}{2}\frac{|\Gamma|}{\Gamma}\sin[2|\Gamma|.t] & 0 & \frac{-\Gamma^*}{\Gamma}\frac{e^{-2B}}{2}(1-\cos[2|\Gamma|.t]) & \frac{e^{-2B}}{2}(1+\cos[2|\Gamma|.t]) \end{pmatrix}\begin{pmatrix}\rho_{11}(0,T)\\\rho_{22}(0,T)\\\rho_{33}(0,T)\\\rho_{44}(0,T)\\\rho_{23}(0,T)\\\rho_{32}(0,T)\end{pmatrix}$$

(23)

where
$$B=\frac{\gamma_0}{2}t+\frac{\gamma_0}{2\gamma}(e^{-\gamma t}-1)$$

$$\Gamma = 2(iJ - D_z)$$

$$|\Gamma| = \sqrt{\Gamma\Gamma^*}$$

$\rho_{ij}(0,T)$ is the component of density matrix which is defined in Eq. (5). As we can see, due to our initial condition on baths, the thermal behavior of the system is originated from $\rho_s(0,T)$.

Finally, before the end of this section, it should be stated that, there are several ways to find solutions to open quantum systems. But sometimes it is difficult to get an analytical answer and we have to use the appropriate approximation. One of these methods is using the Kraus operators to find time evolution of the system after tracing over on the environment part. However, the use of unitary evolution on "system + environment" and some approximations can be useful in obtaining analytical and qualitative responses. It should be noted that, although the solutions obtained by the approximate methods are qualitative, there are other methods such as fractional derivatives and super statistical methods to approximate the solutions to reality. In this paper, we obtain the density matrix of system with initial temperature T, using these standard approximation methods.

The study of entanglement behavior in open quantum systems that involves DM interaction is of particular importance for the following reasons.
   (1) All quantum gates are based on the entanglement between qubits.
   (2) DM interaction is reported in Semiconductors quantum dot [29]. Such materials will be used to build quantum gates.

## 3. specific heat of an open quantum system

In this section, we present a formula to calculate the specific heat of an open quantum system using its density matrix. Two scenarios for obtaining specific heat are reported in papers [30-32]. The first one uses the measurement of the kinetic energy of the free particles, while the second is based on the reduced partition function. We use the second one, namely, a definition for partition function of the system is based on the ratio of the partition functions of the total system and the heat bath, which traditionally is identified as the partition function of an open quantum system:

$$Z_s = \frac{Tr_{sb}[\exp(-\beta H)]}{Tr_b[\exp(-\beta H_B)]} \quad (24)$$

where H and $H_b$ are defined in Eq. (1), and Eq. (3), respectively. The specific heat is defined as (see the Appendix),

$$C = k_B \beta^2 \frac{\partial^2}{\partial \beta^2} Ln(Z) \quad (25)$$

The specific heat of principle system is the difference of the specific heat of total system and baths.

$$C_s(t,T) = C_{sb}(t,T) - C_b(t,T) \quad (26)$$

The thermal density matrix at time $t$ can be obtained as

$$\rho(t,T) = \frac{1}{Z(t,T)} \exp[-\beta H] \quad (27)$$

Using the above equations we can get

$$C_s(t,T) = -4\, k_B \beta^2 \frac{\partial^2}{\partial \beta^2} ( Tr[\, Ln(\rho_{sb})\,] - Tr_b[\, Ln(\rho_b)])(28)$$

After a little algebra and applying some approximations, it can be shown that (see Appendix):

$$C_s(t,T) = -4\, nk_B \beta^2 \frac{\partial^2}{\partial \beta^2} Tr\left[Ln\left(\rho_s^{diag}(t,T)\right)\right] \quad (29)$$

Since the diagonal elements of $\rho_s^{diag}(t,T)$ are the eigenvalues of it, one can write

$$C_s(t,T) = -4\, n\, k_B \beta^2 \frac{\partial^2}{\partial \beta^2} \sum_{i=1}^{m} Ln(\eta_i). \quad (30)$$

where $m\ (=4)$ and $\eta_i$ are the dimension of the Hilbert space and the eigenvalues of density matrix of the system, respectively. $n$ is the total number of modes for the bath. $k_B$ is the Boltzmann constant and $\beta = \frac{1}{k_B T}$. The eigenvalues of $\rho_s(t,T)$ that are the diagonal entries of $\rho_s^{diag}(t,T)$ have the following form

$$\eta_1 = \frac{1}{2}(\,(\rho_{11} + \rho_{44}) + |\rho_{11} - \rho_{44}|\,) \quad (30a)$$

$$\eta_2 = \frac{1}{2}(\,(\rho_{11} + \rho_{44}) - |\rho_{11} - \rho_{44}|\,) \quad (30b)$$

$$\eta_3 = \frac{1}{2}(\,(\rho_{22} + \rho_{33}) + \sqrt{(\rho_{22} - \rho_{33})^2 + 4|\rho_{23}|^2}\,) \quad (30c)$$

$$\eta_4 = \frac{1}{2}(\,(\rho_{22} + \rho_{33}) - \sqrt{(\rho_{22} - \rho_{33})^2 + 4|\rho_{23}|^2}\,) \quad (30d)$$

We define $C_s^{(n)}(t,T) = \frac{C_s(t,T)}{4nk_B}$ and plot it as a function of time and $k_B T$. It should be mention that $T$ is the initial temperature of the system. In Figs. 1(a), and 1(b), as we can see, $C_s^{(n)}$ is saturated at big time and decreases for higher temperature. The specific heat based on effective partition function and defined in Eq. (24), can become negative. This behavior can be interpreted as the change in the specific heat of baths when the system degree of freedom is added to it. This negative value appears in initial moments, at the moment of the system attaches to the environment. In later times, the system reaches to a thermodynamic stability with the environment and the negative values cannot be seen in specific heat. In Fig. 2, we plot the $C_s^{(n)}$ as a function of $k_B T$ and coupling constant $J$. Except for $T \to 0K$ (because of, it is impossible by any procedure, no matter how idealized, to reduce the temperature of any system to zero temperature in a finite number of finite operations), $C_s^{(n)}$ is always well-define in terms of time $t$, coupling constant $J$ and initial temperature of the system $T$.

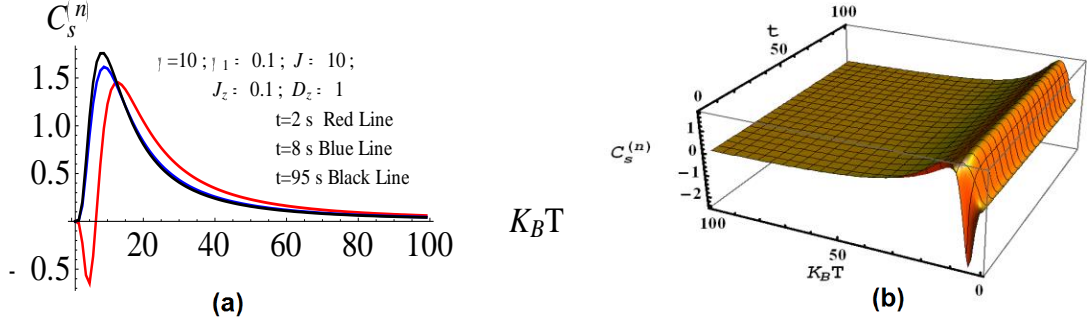

Figure 1. (a) $C_s^{(n)}$ versus $k_B T$ in different time and (b) 3D-plot of $C_s^{(n)}$ versus $k_B T$ and time.

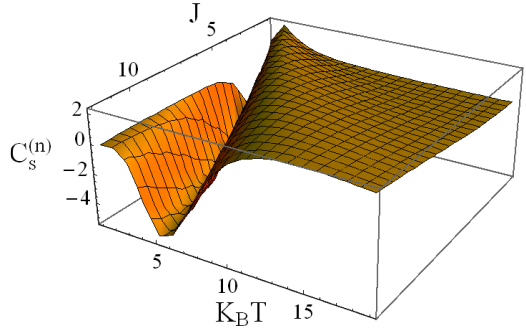

Figure 2. $C_s^{(n)}$ as a function of temperature and J at t= 0.5 s

Now we are looking for a connection between the entanglement of the system and the specific heat. Before that, in order to quantify the entanglement of formation, we use the following equation [33-34].

$$EOF = h\left(\frac{1+\sqrt{1-C^2}}{2}\right) \quad (31)$$

where $h(x) = -x \, Log_2 x - (1-x) Log_2(1-x)$ and $C = \max\{0, \lambda_1 - \lambda_2 - \lambda_3 - \lambda_4\}$ is the concurrence and $\lambda_i (i = 1,2,3,4)$, are the square roots of the eigenvalues of the operator $\rho_s(t,T)\widetilde{\rho_s(t,T)}$ in descending order with $\lambda_1 \geq \lambda_2 \geq \lambda_3 \geq \lambda_4$. $\widetilde{\rho}_s(t,T)$ is defined as

$$\widetilde{\rho}_s(t,T) = (\sigma_{1y} \otimes \sigma_{2y})\rho_s^*(t,T)\,(\sigma_{1y} \otimes \sigma_{2y})$$

Here $\rho_s(t,T)$ is the density matrix of the system for the DM model which has the X- type form. $\sigma_{1y}$ and $\sigma_{2y}$ are the normal Pauli operators. In Fig. 3, we have depicted the EOF as a function of time and coupling constant $J_i$. As we can see, the amount of entanglement decrease with time and increases with J. The notable effect is the existence of successive sudden deaths in the entanglement for larger J. This indicates that the EOF curve in terms of time and $J_i$ is not stable and unpredictable. To realize the importance of this instability, we know that quantum gates work based on entangled states [18]. So, in these areas, we do not have a stable quantum gate (basically, the reason for paying attention to non-Markovian processes is the loss of entanglement with time). In Fig. 4, we can see the behavior of EOF vs. time, $D_z$ and $K_B T$. As can be seen, in larger values of $D_z$, the EOF with drastic changes eventually becomes zero. This shows us choosing the systems with the small spin-orbit coupling is important to have stable entanglement. Also, according to our results, for small J, when the temperature increases, the EOF reduces to zero monotonically. This means that in zero temperature and at any time, the system possesses maximum entanglement. Therefore, the specific heat cannot be diverge in absolute zero temperature [6].

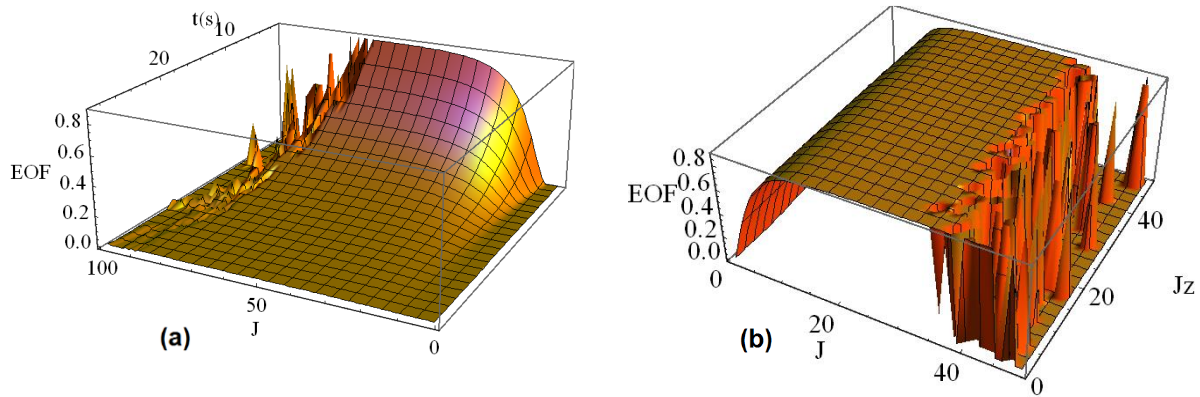

Figure 3. The plot of EOF at $\beta = 0.1$ as a function of (a) Coupling constant J and time (b) Coupling constant J and $J_z$. As can be seen the entanglement of formation is maximum for all $J_z$ and for large amount of J.

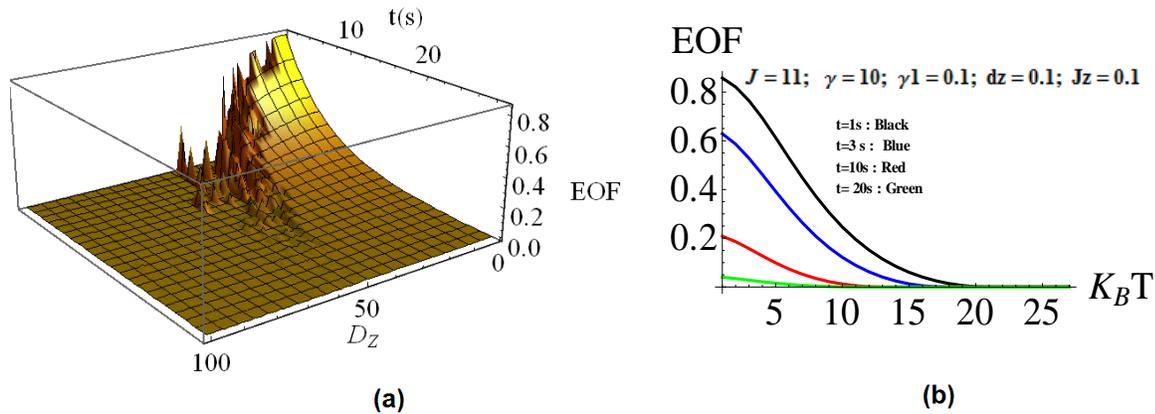

Figure 4. (a) The plot of EOF (for $\beta = 0.1$) as a function of Spin-Orbit coupling $D_z$ and time. In small $D_z$, the EOF is maximum and decrease with time monotonically. (b) The plot of EOF as a function of temperature in different time. We can see in zero temperature, the entanglement is maximum.

Now, before the end of this section, we'll look at some of other results in this paper. As shown in Figures (3) and (4), the interaction parameters of DM influence the entanglement behavior. In some of these parameters, the entanglement exhibits an unpredictable behavior, which makes it difficult to create any quantum gate. These Figures can provide a suitable address (based on $J_z$, $J$ and $D_z$) in selecting materials for solid gates. In addition, we obtained,

(1) A formula for calculating the specific heat of an open quantum system using its density matrix. The method presented here seems to be simpler than other methods, like the Area Law.
(2) At lower initial temperatures, the amount of entanglement of formation in systems that involve the interaction of DM and exposed to the environment is greater (Fig.4b).
(3) As we know, in Hamiltonian of the system (Eq.2), the $D_z$ is a coefficient that shows the spin disordering. This coefficient is the source of weak ferromagnetic property in paramagnetic materials. In accordance with Fig.4a, magnetic disorder reduces entanglement. In large amounts of $D_z$ and spin disorder, sudden deaths can be observed in the entanglement.
(4) In this paper, the case $J > 0$ (which is the symbol of the spin pair coupling and antiferromagnetic property of the system) has been studied. As shown in Fig. 3, with increasing J, the entanglement is rapidly saturated. According to Eq. 5, this is due to the exponential increase of the coefficients u and v as a function of J. The rapid increase of these coefficients causes the spin anti isotropic in the initial thermal state. The result is an increase in entanglement.
(5) As shown in Fig. 4, the entanglement decreases with increasing initial temperature of the system. This is due to the initial thermal fluctuations of the system.

Now, let's take a look at the example below to understand the effects of DM and environment on the performance of the universal swap gate.

**Example.** The effect of DM interaction and environment on the performance of the Swap gate

As we know, Swap Gate is made with two qubits and in the basis set $\{|00\rangle, |10\rangle, |01\rangle, |11\rangle\}$ (in the absence of DM interaction and environmental effects), is shown with the following Unitary operator.

$$U_{Swap} = \begin{pmatrix} 1 & 0 & 0 & 0 \\ 0 & 0 & 1 & 0 \\ 0 & 1 & 0 & 0 \\ 0 & 0 & 0 & 1 \end{pmatrix}. \qquad (31)$$

Under the operation of this operator, the basis are transformed as follows.

$$U_{Swap}|00\rangle = |00\rangle, U_{Swap}|11\rangle = |11\rangle, U_{Swap}|01\rangle = |10\rangle, U_{Swap}|10\rangle = |01\rangle. \qquad (32)$$

However, the result of its operation in the presence of DM is obtained as follows [35].

$$U_{Swap}^{DM}|00\rangle = |00\rangle, U_{Swap}^{DM}|11\rangle = |11\rangle, U_{Swap}^{DM}|10\rangle = e^{\frac{i}{2}(\theta_1 - 2\theta_D)}|01\rangle, U_{Swap}^{DM}|10\rangle = e^{\frac{i}{2}(\theta_1 + 2\theta_D)}|01\rangle. \qquad (33)$$

Where

$$\theta_1 = \frac{1 - \pi\sqrt{1+D^2}}{\sqrt{1+D^2}}, \quad \theta_D = Arctg(D), \quad D = \frac{D_z}{J}. \qquad (34)$$

Now imagine that a system consisting of two qubits interacting with DM and exposed to the bosonic environment, has the following initial state.

$$\rho(0) = |01\rangle\langle 01|. \qquad (35)$$

We know that the matrix elements evolve according to eq. (23). The density matrix after the time evolution can be obtained as follows.

$$\rho(t) = \frac{e^{-2B}}{2}(1 - \cos[2\,|\Gamma|.t])|01\rangle\langle 01| + \frac{e^{-2B}}{2}(1 + \cos[2\,|\Gamma|.t])|10\rangle\langle 10| + (1 - e^{-2B})|11\rangle\langle 11| + \frac{e^{-2B}}{2}\frac{|\Gamma|}{\Gamma^*}\sin[2|\Gamma|.t]|10\rangle\langle 01| + \frac{e^{-2B}}{2}\frac{|\Gamma|}{\Gamma}\sin[2|\Gamma|.t]|01\rangle\langle 01|. \qquad (36)$$

Now imagine that Gate switches at time t.

$$\rho'(t) = U_{Swap}^{DM}.\rho(t).U_{Swap}^{DM\,\dagger} \qquad (37)$$

$$\rho'(t) = \frac{e^{-2B}}{2}(1 - \cos[2\,|\Gamma|.t])|10\rangle\langle 10| + \frac{e^{-2B}}{2}(1 + \cos[2\,|\Gamma|.t])|01\rangle\langle 01| + (1 - e^{-2B})|11\rangle\langle 11| + \frac{e^{-2B}}{2}\frac{|\Gamma|}{\Gamma^*}\sin[2|\Gamma|.t]\,e^{-2i\theta_D}\,|01\rangle\langle 10| + \frac{e^{-2B}}{2}\frac{|\Gamma|}{\Gamma}\sin[2|\Gamma|.t]\,e^{2i\theta_D}\,|10\rangle\langle 01|. \qquad (38)$$

One can see the dependence of $\rho'(t)$ on the $J$ and $D_z$. However, if the gate is switched on in the absence of DM and the environment, the following result is obtained.

$$\rho'(0) = U_{Swap}.\rho(0).U_{Swap}^{\dagger} = |10\rangle\langle 10|. \qquad (39)$$

By comparing the results of equations (8) and (9), it can be seen that the result of the gate operation differs from its ideal and depends on the time and parameters of the system.

## 4. Summary

We considered to a system which is modeled by Heisenberg XXZ chain including DM interaction in which the system is exposed to environment. The density matrix of the system is obtained via Non-Markovian processes. It is observed that for small J and $D_z$, the entanglement decreases monotonically with time. But, for larger $J(>7)\ and\ D_z(>5)$, successive sudden deaths happened in EOF. In these regions of J and $D_z$, the entanglement is not controllable. We presented a formula to calculate specific heat using density matrix for open quantum system. Also, in the obtained graphs, we can see the negative values for the specific heat at low temperatures. The negative values occur when the system is connected to the environment. All these results may be useful to manufacture controllable solid quantum gates (For example, fabrication of graphene-based or GaAs-based gates).

## Appendix

First, we give a proof of the zeroing of the first part of Eq. (8), then we give some useful relations to derive Eq. (25) and finally present the proof of Eq. (28) which is a formula to calculate the specific heat using density matrix of an open quantum system.
   1. In Eq. (8) we can see that:

$$tr_b\{[H_{sb}^I(t), \rho_{sb}^I(0)]\} = 0$$

Using the weak interaction approximation ($\rho^I_{sb}(t) = \rho^I_s(t) \otimes \rho^I_b(0)$) and Eq. (13), it can be proved. The proof is:

$$tr_b\{[H^I_{sb}(t), \rho^I_{sb}(0)]\} = \sum_{i=1}^{2} \sigma_i^+ \rho_s(0) \otimes \sum_j \sum_{n_i} g_{n_i} e^{-i\omega_{n_i^j} t} \langle n_i^j | b_{n_i^j} \rho_b(0) | n_i^j \rangle - \sum_{i=1}^{2} \sigma_i^+ \rho_s(0) \otimes \sum_j \sum_{n_i} g_{n_i} e^{i\omega_{n_i^j} t} \langle n_i^j | \rho_b(0) b_{n_i^j} | n_i^j \rangle + \text{h.c.} \quad \text{(A.1.1)}$$

We know that $\rho_b(0)$ is diagonal in its basis

$$\rho_b(0) = \sum_p \rho_b^p(0) |n_i^p\rangle\langle n_i^p| \quad \text{(A.1.2)}$$

We can see in first term

$$\langle n_i^j | b_{n_i^j} \rho_b(0) | n_i^j \rangle = \sum_p \rho_b^p(0) \langle n_i^j | b_{n_i^j} | n_i^p \rangle \langle n_i^p | n_i^j \rangle = \sum_p \rho_b^p(0) \langle n_i^j | b_{n_i^j} | n_i^p \rangle \delta_{jp}$$

$$= \rho_b^j(0) \langle n_i^j | b_{n_i^j} | n_i^j \rangle = 0 \quad \text{(A.1.3)}$$

With the same manner, all other three terms are neglected.

2. To derive Eq. (25), one can use the following relations.

The thermodynamic energy: $\quad \langle E \rangle = -\frac{\partial \ln(Z)}{\partial \beta} \quad \text{(A.2.1)}$

The specific Heat: $\quad C = \left(\frac{\partial \langle E \rangle}{\partial T}\right)_V \quad \text{(A.2.2)}$

The average energy: $\quad (\Delta E)^2 = \langle E^2 \rangle - \langle E \rangle^2 = k_B T^2 C \quad \text{(A.2.3)}$

3. The Proof for Eq. (28)

We start from the density matrix which is defined as

$$\rho(t, T) = \frac{1}{Z(t,T)} \exp[-\beta H] \quad \text{(A.3.1)}$$

It can be written that

$$Tr[Ln(\rho)] = Tr[Ln(\frac{1}{Z})] - \beta \sum_n E_n \quad \text{(A.3.2)}$$

As we know, the Hamiltonian H and its eigenvalues are not explicitly dependent on β and it can be shown that

$$C_s(t,T) = -4\,k_B\beta^2\,\frac{\partial^2}{\partial\beta^2}(Tr[\,Ln(\rho_{sb})\,] - Tr_b[Ln(\rho_b)]) \quad (A.3.3)$$

Let us now introduce the following relations

$$Tr[A \otimes B] = Tr[A]\,Tr[B] \quad (A.3.4)$$

$$Ln(A \otimes B) = \ln(A^{diag}) \otimes I_n + I_m \otimes \ln(B^{diag}) \quad (A.3.5)$$
$$Tr[\,Ln(A \otimes B)\,] = nTr[\,Ln(A^{diag})\,] + mTr[\,Ln(B^{diag})\,] \quad (A.3.6)$$

where A and B are the positive Hermitian matrixes with dimensions m and n, respectively, and $A^{diag}$ and $B^{diag}$ represent their associated diagonal forms. In the weak interaction approximation between the system and environment, we can write

$$\rho_{sb}(t,T) = \rho_s(t,T) \otimes \rho_b(0) \quad (A.3.7)$$

By applying the above approximation and using the equations (A.3.4)-(A.3.6), it is easy to show that

$$Tr_{sb}[Ln(\rho_{sb})] = Tr\left[Ln\left(\rho_s^{diag}(t,T)\right)\right].Tr[I_n] + Tr[I_4].Tr[Ln(\rho_b^{diag}(0))] \quad (A.3.8)$$

In addition, choosing the initial density matrix as

$$\rho_b(0) = \rho_b^{diag}(0) = \left|N;\ldots,n_j,\ldots,n_p,\ldots\right\rangle_b\left\langle N;\ldots,n_j,\ldots,n_p,\ldots\right| \quad (A.3.9)$$

It is found that:

$$Tr[Ln(\rho_b^{diag}(0))] = 0 \quad (A.3.10)$$

Finally, Eq. (A.3.3) can be rewritten in following form.

$$C_s(t,T) = -4\,nk_B\beta^2\,\frac{\partial^2}{\partial\beta^2}Tr\left[Ln\left(\rho_s^{diag}(t,T)\right)\right] \quad (A.3.11)$$

**ACKNOWLEDGMENTS**

Acknowledge support from their respective budget, Azad Islamic university, Ahvaz Branch, Ahvaz. We thank the anonymous referee whose suggestions have contributed toward the improvement of this Report.